# The Homogeneity Indicator of Learners in Project-based Learning

Samir Bennani[1], Mohammed Khalidi Idrissi[2], Faddouli Nourredine[3], and Yassine Benjelloun Touimi[4]

LRIE-Research Laboratory in Computer Science and Education, Computer Science Department, Mohammadia Engineering School, Mohammed V-Agdal University, Avenue Ibnsina B.P.765 Rabat, Morrocco

**Abstract**
The development of ICT has emerged a new way of learning using electronic platforms: E-learning. In addition, pedagogical approaches have been adopted in teaching based on group learning, such as the project-based teaching. The project-based teaching is an active learning method, based on group work to develop skills and acquire knowledge.
However, the group of students is facing several challenges throughout the project, such as the decision-making group. The group decision generates convergences and divergences among members. Our approach in this article relates to the calculation of the homogeneity of a group of learners during decision making in an educational project.

## 1. Introduction

The As With technological development, new forms of group work have emerged, particularly in the field of distance education (e-learning).
E-Learning has experienced a lot of changes, and improved teaching conditions, while facing the temporal and spatial constraints.
Most learning platforms are more interested in content management, rather than the distance education process. This problem is highlighted in the socio-constructivist theory, which promotes team work and knowledge sharing. Among the socio constructivist models, we focus on the pedagogy project [1]. The Project-based teaching is characterized by collaborative learning, social nature, which promotes negotiation, critical of others, and group decision-making.
Throughout the project, students are confronted with situations of collaborative decision-making to solve a problem. The decision making in an educational project, occurs in all stages of the project: project selection, equipment selection, planning, scenario, and assessment.
The tutor assesses the effectiveness of decision-making, by means of homogeneity indicator. The measure of group homogeneity is used in formative and summative assessment of learners.
In this paper, we propose the calculation of the homogeneity indicator for a group of learners in an educational project.
In the first section we will discuss a state of the art of project-based teaching, and collective decision-making in a collaborative learning situation. Thereafter in the second section, we will study the contribution of the AHP method for collaborative decision making. Then we will define a measure of the group homogeneity, to assess the efficiency of the learner's decisions. The final section will highlight the work in progress and our main perspectives.

## 2. The Project-Based Learning

Project-based teaching is a learning approach which presents some aspects of sustainable learning skills, such as group work, communication, critical thinking, and decision making [2].
This method of learning develops disciplinary and transversal skills of learners. Skills are individual and collaborative type.
The project gives students the opportunity to work in a group for a period of time, as opposed to individual teaching.
A group project exposes students to other's points of view, from which they can learn and accomplish their tasks conveniently. The Group projects provide the opportunity for the development of interpersonal skills, and teamwork, such as communication, planning and time management skills very researched by graduates in the workplace.
Indeed, a group project is considered as a learning process composed of a set of sub processes. The online assessment is a fundamental process of distance learning.
The evaluation process is based on the calculation of a set of collaborative and individual indicators. Indicators are used in all modes of evaluation: prerequisites, formative, and summative [3]. Among these indicators, there is the indicator of homogeneity.

Indeed, students are brought to take collective decisions in all stages of the project, to find effective solutions to problems.

The collaborative decision-making requires the commitment of all members of the group as well the learners express preferences for different solutions by assigning values.

The Analytic Hierarchy Process [20] is a method of multi-criteria decision making, based on the aggregation of individual preferences into collective decision.

However, a decision is considered efficient, if all group members converge towards the same decision. In this case we consider that the group is homogeneous.

As a result, the homogeneity indicator assesses the effectiveness of collaborative decisions during a project. In formative assessment, the homogeneity indicator is calculated for each collaborative decision making, and is made every step of the project.

At the end of the project, the aggregation of the values of the homogeneity indicator gives us a summative evaluation of the homogeneity of the group.

## 3. The decision making in a project-based learning

In a pedagogical project, the process of decision making is a fundamental process in a learner path. Decision making in a group, is made by means of consensus, vote, compromise, geometric mean.

The AHP method [20] is a widespread method that treats the collective decision-making. The AHP method provides a mechanism for expressing the preferences and goals of the participants, and to generate a solution that takes into account individual participant evaluations.

This method develops communication and understanding between decision makers, which opens the way for convergence of preferences, and builds a consensus so that a solution of minimal conflict is generated [14].

In a project, the collaborative decision-making between learners, involves a set of alternative solutions, a set of evaluation criteria, and a group of learners makers.

Alternative decision representing the decisions of learners noted Ai for i = 1... m. The alternatives are evaluated based on a set criteria, Cj, j = 1,2, ..., n. Learners form a group of decision makers noted by Dk, with k the number of individuals involved in the process of collaborative decision-making, with k = 1,2, ..., q.

In theory, solving the multi criteria decision-making MCDA [21], is based on the aggregation of individual solutions by assigning different weights to the evaluation criteria.

Each learner solves the problem of decision individually to get a set of individual solutions, and in the second stage the individual solutions are aggregated using the rules of collective choice, or an algorithm to obtain a group solution.

In the case of educational project, learners are confronted to problem solving, so they gives their opinions, discusses and criticize their peers.

Alternatives are assigned values by the members individually, and in groups.

At start of project, learners are organized to discuss the conception of the project. Then they make the choice of material, the E-learning platform, and documents to be consulted.

The learning scenario is proposed by the tutor, which assigns educational activities (courses, exercise, quiz, etc ...) in order to solve the problems of the project.

Learners collaborate and decide on the delegation and orientation of activities, and the choice of appropriate solutions.

Each group decides on the strategy for performing the necessary tasks of the project, either individually or collaboratively. In our context, we will study the decision of the group, to choose the optimized solution to a problem (Fig.1).

In all stages, the tutor provides the group a set of tasks to execute. Learners carry out assigned tasks, according to an advance planning. Then the tutor performs an evaluation the group's work.

The tutor provides a set of solutions, and learners give preference values for each solution. The AHP process [20], allows aggregating the priorities of solutions to provide a collaborative solution of the group. This collaborative decision is only efficient if the group find a consensus among its members.

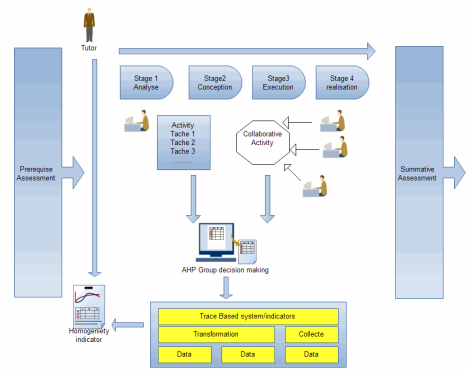

Fig. 1 Homogeneity measure in project-based learning

# 4. The group homogeneity in project-based learning

The Consensus or homogeneity [10] is an opinion or position reached by a group of individuals acting as a whole, generally considered an agreement. Hence, dissent is the complement of consensus.

In a consensus, individuals who want to react by actions want to hear the opponent views, and not impose a decision, and follow conversations that will benefit everyone.

Indicators [6] play a key role in any learning process, including in an e-learning system. The calculation of indicators allows measuring the success of the educational project, by comparing the reached values with those listed at the beginning of the project.

In our case study, we chose to study the homogeneity [7] [10] within a project group indicator during a decision-making. The decision-making of a group of learners used to find an optimized solution. The measurement of the indicator is based on the information theory [8], and known as the Shannon entropy.

## 4.1 Measuring of the homogeneity in collaborative decision making in the project-based learning

At each stage of the project, the distances between the values assigned by the learners, and the values of collective decisions, constitute a variable for measuring the homogeneity of the group.

The preference values are transformed into ranks on a scale of likert [9]. The likert scale [9] is a one-dimensional ordinal scale composed of ordinal values used to collect data by means of categories. The type of data collected frequently involves the determination of the attitude or feelings according to the attributes.

The likert scale [9] is expressed like statements with categories of choice classified from the value 'agreement extreme' to 'extreme disagreement'. The choice of the value on the Likert scale [9] by the participant must be on a single category.

The likert scale [9] can be represented by different numbers of categories, usually two to nine categories are used to convert the subjective opinions to ordinal values [12].

Table 1: Margin specifications

| Ranks Learners | Alternatives | | | | |
|---|---|---|---|---|---|
| | X1 | X2 | X3 | X4 | X5 |
| d1 | R11 | R12 | R13 | R14 | R15 |
| d2 | R21 | R22 | R23 | R24 | R25 |
| d3 | R31 | R32 | R33 | R34 | R35 |
| d4 | R41 | R42 | R43 | R44 | R45 |
| d5 | R51 | R52 | R53 | R54 | R55 |
| Group | Rg1 | Rg2 | Rg3 | Rg4 | Rg5 |

Table 1 shows the ranks of classification of individual and collective preferences. Then we proceed to calculate the distances between the individual ranks of the alternative values, and the ranks of the values assigned by the group.

The total consensus means that the ranking of alternatives by individuals is the same as the ranking of alternatives in the collective group solution

For each individual, there is a set of row: $[R_1^k, R_2^k, ....., R_m^k]$ with $R_i^k$ is the rank of alternative i, according to the decision maker preferences k. The alternatives with the same score are assigned the same position.

We use the concept of Shannon entropy [8] to quantify the distribution of the difference between the decision of learner and group.

The formula of the Shannon entropy is given by:

$$H(X) = -\sum_{i=1}^{n} p(x_i) \log_2 p(x_i) \quad (1)$$

One of the attributes of the entropy equation is its ability to measure the degree of uncertainty in the sampling process. In the biological field [17], the calculation of the entropy of a community with s species informs us about the uncertainty of the identity of a species in a sample, but not on the number of species.

The Shannon entropy [8] in the case of a multi-criteria decision is defined for each decider k:

$$H^k = -\sum_{i=1}^{n} p_i^k \log_2 p_i^k \quad (2)$$

With i = 1,..,n the number of alternatives, and $p_i^k$ the probability of rank $R_i^k$, calculated by dividing $R_i^k$ the number of on the total number of rank values, for alternative i and for the decision maker k.

The diversity D of the first order for learner k is given by:

$$D^k = \exp(H^k) \quad (3)$$

The uniformity measures the degree of deviation of the individual overall preference, against the overall collective preference:
$$U^k = \frac{H^k}{\log_2(n)} \quad (4)$$

$H^k$: The Shannon entropy for decision-maker k.

$\log_2(n)$: The Napierian logarithm of the number of alternatives n.

In the case where $U^k$ is close to 1, then the distribution $D_i^k$ avec $(1 \leq i \leq n)$, is uniform. Otherwise the distribution $D_i^k$ is not uniform, and $U^k$ is close to 0.

For a group of k learners the Shannon entropy alpha is the average of the individual entropy.

$$H_\alpha = -\sum_{j=1}^{k} w_j \sum_{i=1}^{N} p_{ik} \ln(p_{ik}) \quad (5)$$

The alpha diversity measures the average distance distribution of preferences for the alternatives for each learner group.

The first order Diversity alpha will be: $D_\alpha = \exp(H_\alpha)$ (6)

The concept of diversity alpha allows the partition of diversity into two independent components alpha and beta [18]:

$$H_\gamma = H_\alpha + H_\beta \quad (7)$$

In the context of a group of learners, i = 1.... k, the concept of Shannon diversity gamma is defined as:

$$H\gamma = -\sum_{j=1}^{K}\left[\sum_{i=1}^{n} w_i \cdot p_{ij}\right] \cdot \left[\ln(\sum_{i=1}^{n} w_i p_{ij})\right] \quad (8)$$

The first order diversity gamma: $D_\gamma = \exp H_\gamma$ (9)

So we can deduce the beta diversity of the first order:

$$D_\beta = D_\alpha / D_\gamma \quad (10)$$

The first order beta diversity is a measure of variations in distances between group members, and therefore we can deduce the degree of homogeneity of the group.

The inverse of beta diversity is a measure of the homogeneity of the group [19]:

$$M = 1/D_\beta \quad (11)$$

The homogeneity indicator takes a value range between 0 and 1. A value 1 means absolute homogeneity and value 0 complete dissensions. A small variation in beta diversity means high homogeneity and group consensus.

The value of the homogeneity is based on the calculation of the first-order diversity. The diversity of the first order is the exponential of Shannon entropy (Shannon, 1949).

### 4.2 Illustrative example

As an illustration we take values from a study of the choice of urban sites in a construction project [16]. The group is composed of three learners that will express preferences among five alternatives.

The evaluation criteria for site selection are: the distance to the habitats, the distance to hotels, the distance to the main avenue, and the distance to the highway.

The tutor assigns to learners the following activities: measuring distances between sites and habitats, measuring the average distance between sites and the highway, measuring the distance to the main avenue, measuring the distance from the highway, and estimate the cost of construction of urban sites.

After having completed the tasks, each learner assigns a preference value for each site according to the evaluation criteria. Using the AHP method [16], we compute the solution of the group by aggregating the values of the priorities of each solution.

We classify individual preferences and group, according to likert scale [9], and we find the rank of each value of the alternatives in the table 2 below.

Table 2: the ranks of individual preferences and collective

| Learners | ALT1 | ALT2 | ALT3 | ALT4 | ALT5 |
|---|---|---|---|---|---|
| L1 | 2 | 4 | 3 | 1 | 5 |
| L2 | 2 | 3 | 5 | 4 | 1 |
| L3 | 1 | 3 | 4 | 2 | 2 |
| Group Solution | 1 | 3 | 5 | 4 | 2 |

The distance is calculated between the group solution, and preferences of learners through dissimilarity function. The distance $D_i^k$ is the difference between the rank of the alternative i in the group solution, and $R_i^k$ the rank of preference of alternative i for decision maker k, and we find the results in Table 3.

Table 3: Distance between the preferences of learners and the group

|  | DISTANCE | | | | |
|---|---|---|---|---|---|
| Learner | ALT1 | ALT2 | ALT3 | ALT4 | ALT5 |
| L1 | 1 | 1 | 2 | 3 | 3 |
| L2 | 1 | 0 | 0 | 0 | 1 |
| L3 | 0 | 0 | 1 | 2 | 0 |
| Sum of the distance | 2 | 4 | 6 | 9 | 5 |

By dividing the values of each rank by the sum of the distance, it results the distribution of alternatives preferences (Table 4).

Table 4: Probability of distance distributions

| Probabilities | | | | | |
|---|---|---|---|---|---|
| Learners | ALT1 | ALT2 | ALT3 | ALT4 | ALT5 |
| L1 | 0,500 | 0,250 | 0,333 | 0,333 | 0,600 |
| L2 | 0,500 | 0,000 | 0,000 | 0,000 | 0,200 |
| L3 | 0,000 | 0,000 | 0,167 | 0,222 | 0,000 |

By using equation (1) (2) and (3) we calculate the alpha diversity index for each learner j = 1,..., k as shown in Table 5.

Table 5: The Indices of alpha diversity and uniformity of learners

| Learners | Alpha Index | Alpha diversity | Uniformity |
|---|---|---|---|
| L1 | 1,732 | 5,652 | 1,076 |
| L2 | 0,668 | 1,951 | 0,415 |
| L3 | 0,633 | 1,883 | 0,393 |

The learner L1 shows an alpha index and alpha diversity, superior to other learners L2 and L3, therefore a distribution probability, more uniform than the others.

Regarding the learner group, equation (4) is used to measure the index of alpha diversity of the group, and equation (5) to measure the diversity index gamma.

The difference between gamma and alpha diversity gives us the value of beta diversity between group members, and therefore the value of the indicator of homogeneity.

Table 6: calculating the homogeneity indicator of the group

| Group alpha index | 0,927 |
|---|---|
| Group gamma index | 1,508 |
| Group bêta index | 0,581 |
| Group bêta diversity | 1,787 |
| Group homogeneity | 0,560 |

Note that the homogeneity of the group is average (Table 6), so we can calculate the homogeneity between group members to detect disagreements between members.
The homogeneity matrix resulting of calculating the homogeneity among the members of the group of learners is shown in Table 7.

Table 7 the indicator of homogeneity among members of the group of learners

| The homogeneity degree | L1 | L2 | L3 |
|---|---|---|---|
| L1 | 1,000 | 0,691 | 0,361 |
| L2 | 0,691 | 1,000 | 0,686 |
| L3 | 0,361 | 0,686 | 1,000 |

According to the matrix, the peers (L1, L2) and (L2, L3) are consistent, while (L1, L3) show little homogeneity. Therefore, the tutor proposes new alternatives to promote the homogeneity of the group.

## 5. Measuring the Summative Homogeneity of the Project- Based Learning

The indicator of homogeneity of the project is calculated by aggregating the values of the indicator of homogeneity throughout the project.
At the beginning of the project, a prerequisite test administrated to learners to calculate a threshold value of the indicator $H_{seuil}$.

$$H_{tot} = \sum_{m=1}^{p} H_m$$

The average value of the indicator m= 1 ... p, the number of values of the indicator of homogeneity in the educational project.
Beyond this threshold, it is considered that the group is homogeneous, otherwise the group is heterogeneous.

## 6. Conclusion & perspectives

In this article we proposed a homogeneity indicator for a group of learners in an educational project. The calculation of this indicator aims, the assessment of a group of learners in their learning path.

The control and monitoring of learners in stages of the project serve to regulate the path of learners by developing new activities.

At the end of the project, the aggregation of values for this indicator gives a value of the homogeneity degree achieved. The summative evaluation of the project consists in comparing the value of indicator, with the value mentioned in the beginning of the project. Therefore the success of the educational project is measured.

However this work is limited to the calculation of the homogeneity in the process of collaborative decision-making, treated with AHP method.

Into perspective of this work, we will define a global formula for the calculation of the indicator of homogeneity not only in the decision-making situations.

**Bennani Samir :** Engineer degree in Computer Science in 1982; Doctorate degree in Computer Science, PhD in Computer Science in 2005; Former chief of the Computer Science Department at the Mohammadia School of Engineers (EMI); Professor at the Computer Science Department-EMI; 15 recent publications papers between 2008 and 2013; Ongoing research interests: SI, Modeling in Software Engineering, Information System, E-Learning, content engineering, tutoring, assessment and tracking.

**Mohammed Khalidi Idrissi** Doctorate degree in Computer Science in 1999; Assistant Professor at the Computer Science Department at the Mohammadia School of Engineers (EMI); 10 recent publications papers between 2007 and 2013; Ongoing research interests: Web services, elearning, evaluation and information systems.

**Nourrdine Faddouli** Doctorate degree in Computer Science in 1999; Assistant Professor at the Computer Science Department at the Mohammadia School of Engineers (EMI); 10 recent publications papers between 2007 and 2013; Ongoing research interests: Web services, elearning, evaluation and information systems.

**Benjelloun Touimi Yassine** Engineer degree in Computer Science in 2003; PhD Student in Computer Science; engineer a defence department of morocco ; 6 recent publications between 2010 and 2013 ; Ongoing research interests: E-learning and Assessment; data mining.

.